\documentclass[superscriptaddress,secnumarabic,nobibnotes,aps,prd,showkeys,showpacs,nofootinbib,onecolumn]{revtex4}
\usepackage{graphicx}
\usepackage{epsf}
\usepackage{amsmath}
\usepackage{amsfonts}
\usepackage{amssymb}
\usepackage{epstopdf}
\usepackage{hyperref}

\input{tcilatex}

\begin{document}

\title{Stability of the Kasner Universe in $f(T)$ Gravity}
\author{Andronikos Paliathanasis}
\email{anpaliat@phys.uoa.gr}
\affiliation{Instituto de Ciencias F\'{\i}sicas y Matem\'{a}ticas, Universidad Austral de
Chile, Valdivia 5090000, Chile}
\affiliation{Institute of Systems Science, Durban University of Technology, PO Box 1334,
Durban 4000, Republic of South Africa}
\author{Jackson Levi Said}
\email{jackson.said@um.edu.mt}
\affiliation{Institute of Space Sciences and Astronomy, University of Malta, Msida, MSD
2080, Malta}
\affiliation{Department of Physics, University of Malta, Msida, MSD 2080, Malta}
\author{John D. Barrow}
\email{J.D.Barrow@damtp.cam.ac.uk}
\affiliation{DAMTP, Centre for Mathematical Sciences, University of Cambridge,
Wilberforce Rd., Cambridge CB3 0WA, UK}

\begin{abstract}
$f(T)$ gravity theory offers an alternative context in which to consider
gravitational interactions where torsion, rather than curvature, is the
mechanism by which gravitation is communicated. We investigate the stability
of the Kasner solution with several forms of the arbitrary lagrangian
function examined within the $f(T)$ context. This is a Bianchi type--I
vacuum solution with anisotropic expansion factors. In the $f(T)$ gravity
setting, the solution must conform to a set of conditions in order to
continue to be a vacuum solution of the generalized field equations. With
this solution in hand, the perturbed field equations are determined for
power-law and exponential forms of the $f(T)$ function. We find that the
point which describes the Kasner solution is a saddle point which means that
the singular solution is unstable. However, we find the de Sitter universe
is a late-time attractor. In general relativity, the cosmological constant
drives the the isotropization of the spacetime while in this setting the
extra $f(T)$ contributions now provide this impetus.
\end{abstract}

\keywords{Cosmology; $f(T)$--gravity; Kasner Universe; Stability;}
\pacs{98.80.-k, 95.35.+d, 95.36.+x}
\maketitle
\date{\today}

\section{Introduction}

Alternatives theories of gravity are one of the most direct approaches to
reproducing the observed late--time expansion of the Universe \cite%
{Riess:1998cb,Perlmutter:1998np,clifton}. Moreover, general relativity (GR)
suffers from several consistency \cite{weinberg1972gravitation,Martin:2012bt}
problems that must eventually be tackled. One must then ask what theories of
gravity preserve the favorable parts of GR while resolving some of the
outstanding issues. A similar course was taken with regard to cosmological
models in $f(R)$ gravity in order to see which features of the Friedmann
universes of GR were retained in these larger theories \cite{barott}. One
framework that has gained momentum in recent years is that of teleparallel
gravity (TEGR) \cite{ein28,Tsamp}~and its modification, the $f\left(
T\right) $ teleparallel theory \cite{Ferraro,Lin2010,sar01}. This theory of
gravity provides a different perspective on the mechanism of gravity. In GR
gravity is realized through curvature on spacetime, while in teleparallel
gravity its influence occurs through the connection defined by an
nonholonomic basis. These theories also have non-trivial implications for
Lorentz invariance \cite{SLB,SLB2}.

The Kasner universe \cite{kasner1} is probably the most famous closed-form
cosmological solution of GR in vacuum. In general, the Kasner solution
describes an anisotropic metric in which the space directions are Killing
translations, that is, the spacetime is invariant under a three-dimensional
Abelian translation group. For a four-dimensional spacetime, the Kasner
metric has three parameters, namely the Kasner indices, which must satisfy
the two so-called Kasner algebraic relations, so it is a one-parameter
family of solutions. Specifically, the values of the parameters are defined
on the real number line by the intersection of a three-dimensional sphere of
radius unity, and a plane in which the sum of those parameters is one%
\footnote{%
For higher spacetime dimensions, there are more Kasner exponents and the
Kasner relations are defined in a similar way for those parameters.}. There
are various applications of the Kasner universe; for instance, in
higher-dimensional theories such the Kaluza--Klein theory \cite{DemaretKL}.
Moreover, the evolution of the Mixmaster universe when the effects of the
Ricci scalar of the three-dimensional spatial hypersurface are negligible,
is described by the Kasner solution. More specifically, it has been shown
that power-law models can approximate general Bianchi models at intermediate
stages of their evolutions and at early and late-time asymptotes \cite%
{coleyb}, Because of the simplicity and the importance of the Kasner
solution it has been the subject of study in various modified or
higher-order theories, for instance see refs.\cite%
{kas1,kas2,kas3,kas4,kas5,kas6,kas7,barcl,barcl2,topo,bar01}. It has also
played an important role as a paradigm for the study of the observational
consequences of anisotropic expansion during key periods of cosmological
history involving quantum particle creation \cite{zs}, baryosynthesis \cite%
{bt}, inflation \cite{bher}, massive particle survival \cite{bmass},
magnetic field evolution \cite{silk, skew}, primordial nucleosynthesis \cite%
{HT, jbBBN}, and the temperature isotropy and statistics of the microwave
background \cite{BJS, bher, cmb}.

In alternative theories of gravity it is preferable for GR to be recovered
in some well-defined limit, while the stability of GR solutions in these
theories is a subject of special interest. A detailed analysis on the
existence and the stability of anisotropic solutions in higher-order
theories is performed in ref.\cite{midd}. In particular, Kasner-like
solutions which provide cosmological singularities with inflationary
solutions were determined. However $f\left( T\right) $ gravity is a
second-order theory and very few anisotropic solutions are known in the
literature \cite{vac1,vac2}. Recently anisotropic vacuum solutions of the
Kasner type were found in ref.\cite{ftSing}. Specifically, the conditions
for the $f\left( T\right) $ function in which a nonlinear $f\left( T\right) $
theory provides a solution of TEGR/GR were determined and consequently the
Kasner universe is found to satisfy the field equations in $f\left( T\right) 
$~gravity. In this work, we are interested in the stability of the Kasner
universe. The plan of the paper is as follows.

In Section \ref{field} the formal theory and definitions of the teleparallel
gravity are presented, after which the field equations of $f\left( T\right) $
gravity are given. Moreover, we review some solutions of GR in the nonlinear 
$f\left( T\right) $ theory context. Section \ref{stability} includes the
main material of our analysis. We derive the field equations for the vacuum
Bianchi I universe and show that the Kasner solution satisfies the field
equations. We continue by studying the stability of the trajectories which
describe the Kasner solution for two $f\left( T\right) $ theories of special
interest in the literature. Specifically we consider the power--law $f\left(
T\right) =T+f_{0}T^{n}$ \cite{Ferraro} and the exponential $f\left( T\right)
=T+f_{0}\left( 1-e^{-pT^{n}}\right) $ forms \cite{Lin2010,bamex} (this
covers both forms presented in the literature). From our analysis we find
that the point in the space of the solutions which describes the Kasner
universe is a saddle point which means that the Kasner universe is unstable
in $f\left( T\right) $ gravity.

An analysis of the critical points for the field equations is performed in
Section \ref{deSitter}. We find that Minkowski spacetime and the de Sitter
universe are critical points for the field equations. The point which
describes Minkowski spacetime is a saddle point, while the de Sitter
solution is described by a hyperbolic point when the expansion rate is
negative, and by a sink point which can describe an expanding universe.
Finally, in Section \ref{conc} we discuss our results and draw conclusions.

In this work, Latin indices are used to refer to inertial frames and Greek
indices refer to global coordinates.

\section{$f\left(T\right)$~teleparallel gravity}

\label{field} For the convenience of the reader, we briefly introduce the
basic framework of teleparallel gravity. The existence of a nonholonomic
frame is necessary for the teleparallel gravity and consequently for the $%
f\left( T\right) $ teleparallel gravity. Consider $e_{\phantom{i}\mu }^{i}$
as the description of the nonholonomic frame and $\mathbf{e}=$ $e_{i}\left(
x^{\mu }\right) \partial _{\mu }$ where the commutator~$[e_{\mu },e_{\sigma
}]=c_{(\nu \sigma )}^{...\mu }e_{\mu }$ and $c_{\nu \sigma }^{...\mu }$ are
pure antisymmetric geometric objects, that is, $c_{(\nu \sigma )}^{...\mu
}=0. $\medskip

In general, this leads to the connection being defined as 
\begin{equation}
\Gamma _{\nu \sigma }^{\mu }=\left\{ \overset{\mu }{\nu \sigma }\right\} +%
\frac{1}{2}\gamma ^{\mu \lambda }(c_{\lambda \nu ,\sigma }+c_{\lambda \sigma
,\nu }-c_{\nu \sigma ,\lambda }),  \label{lf.00}
\end{equation}%
where $\left\{ \overset{\mu }{\nu \sigma }\right\} $ denotes the standard
Levi--Civita connection of GR and $c_{\mu \nu \rho }=\gamma _{\lambda \rho
}c_{\mu \nu }^{...\rho }$. By assuming that $\mathbf{e}$ are orthonormal
then they form a vierbein field $\mathbf{e}{(x^{\mu })}$ and $\gamma _{\mu
\nu }$ is the Minkowski metric, $\eta _{\mu \nu }$. Furthermore, from Eq.(%
\ref{lf.00}) it follows that if the connection $\Gamma _{\nu \sigma }^{\mu }$
contains only antisymmetric parts and that it is given by the following
simplified form%
\begin{equation}
\Gamma _{\nu \sigma }^{\mu }=\frac{1}{2}\eta ^{\mu \lambda }(c_{\lambda \nu
,\sigma }+c_{\lambda \sigma ,\nu }-c_{\nu \sigma ,\lambda }),  \label{lf.001}
\end{equation}%
with the property $\Gamma _{\nu \sigma }^{\mu }=-\Gamma _{\sigma \nu }^{\mu
} $.

The latter antisymmetric connection lead to the definition of the torsion
tensor 
\begin{equation}
T_{\phantom{\lambda}\mu \nu }^{\lambda }=e_{a}^{\phantom{a}\lambda }\partial
_{\mu }e_{\phantom{a}\nu }^{a}-e_{a}^{\phantom{a}\lambda }\partial _{\nu }e_{%
\phantom{a}\mu }^{a},  \label{lf.002}
\end{equation}%
while the Riemann tensor vanishes \cite{sar01}. The difference between GR
and teleparallel gravity is characterized through the contorsion tensor
which is defined as 
\begin{equation}
K_{~~~\beta }^{\mu \nu }=-\frac{1}{2}({T^{\mu \nu }}_{\beta }-{T^{\nu \mu }}%
_{\beta }-{T_{\beta }}^{\mu \nu }).  \label{lf.04}
\end{equation}%
Finally, for convenience the superpotential tensor is defined as 
\begin{equation}
{S_{\beta }}^{\mu \nu }=\frac{1}{2}({K^{\mu \nu }}_{\beta }+\delta _{\beta
}^{\mu }{T^{\theta \nu }}_{\theta }-\delta _{\beta }^{\nu }{T^{\theta \mu }}%
_{\theta }).  \label{ft.03}
\end{equation}

This leads to the torsion scalar term 
\begin{equation}
T={S_{\beta }}^{\mu \nu }{T^{\beta }}_{\mu \nu },  \label{ft.02a}
\end{equation}%
which plays the role of lagrangian density in the teleparallel equivalent of
general relativity (TEGR).

With the torsion scalar to hand, we can quantify the difference in
lagrangian densities between GR and TEGR, which is 
\begin{equation}
T=-R+2e^{-1}\partial _{\nu }\left( eT_{\rho }^{~\rho \nu }\right) ,
\label{ft.33}
\end{equation}%
where the extra term acts as a boundary term so that both theories produce
GR at the level of their field equations.

In $f\left( T\right) $ theory, the gravitational action is generalized to an
arbitrary function, that is 
\begin{equation}
S_{f\left( T\right) }=\frac{1}{16\pi G}\int d^{4}xe\left( f(T)\right) +S_{m}
\label{ft.05}
\end{equation}%
in which $e=\det (\mathbf{e})=\sqrt{-g}$, and $g_{\mu \nu }=\eta _{ij}e_{%
\phantom{i}\mu }^{i}e_{\phantom{j}\nu }^{j}$. By definition the invariant $T$
admits only first order derivatives of the vierbeins which means that the
field equations following the lagrangian density in Eq.(\ref{ft.05}) are of
second order. Indeed, the variation with respect to the vierbein provides
that the gravitational field equations \cite{SLB} 
\begin{align}
& e^{-1}\partial _{\mu }(ee_{i}^{\phantom{i}\rho }S_{\rho }{}^{\mu \nu
})f_{T}+e_{i}^{\phantom{i}\lambda }T^{\rho }{}_{\mu \lambda }S_{\rho
}{}^{\nu \mu }f_{T}+  \notag  \label{ft.06} \\
& \ \ \,+e_{i}^{\phantom{i}\rho }S_{\rho }{}^{\mu \nu }\partial _{\mu }({T}%
)f_{TT}+\frac{1}{4}e_{i}^{\phantom{i}\nu }f({T})=4\pi Ge_{i}^{\phantom{i}%
\rho }\mathcal{T}_{\rho }{}^{\nu }.
\end{align}%
In the latter expression the tensor $\mathcal{T}_{\rho }{}^{\nu }$ denotes
the energy--momentum tensor of the matter source~$S_{m}$ while $f_{T}$ and $%
f_{TT}$ denote the first and second derivatives of the function $f(T)$ with
respect to $T$.

\subsection{TEGR in nonlinear $f\left(T\right)$-gravity}

$f\left(T\right)$--gravity is a second--order theory and in the limit of a
linear lagrangian density function, $f_{TT}=0$, we recover GR at the level
of equations, a cosmological constant can also be added. On the other hand,
the general $f(T)$ theory provides structural differences in terms of
properties and observational predictions such as Refs.\cite%
{Iorio2,Bas,Farrugia:2016xcw}. While GR satisfies all small scale tests of
gravity, $f(T)$ gravity does not generally violate these tests.

In ref.\cite{barott}, the conditions for the existence and stability of
solutions to $f\left( R\right) $ gravity are investigated in de Sitter and
Friedmann cosmologies. Recently, this work inspired an analogous analysis in 
$f(T)$ gravity in ref.\cite{ftSing} where new conditions were found for the
current context, i.e. the field equations in Eq.(\ref{ft.06}).

Given the TEGR we can describe the Einstein tensor in teleparallel
quantities as 
\begin{equation}
e_{i}^{\phantom{i}\rho }\mathbf{G=}2\left( e^{-1}\partial _{\mu }(ee_{i}^{%
\phantom{i}\rho }S_{\rho }{}^{\mu \nu })+e_{i}^{\phantom{i}\lambda }T^{\rho
}{}_{\mu \lambda }S_{\rho }{}^{\nu \mu }+\frac{1}{4}e_{i}^{\phantom{i}\nu
}T\right) ,  \label{ft.32}
\end{equation}%
so that the field equations take on the form 
\begin{equation}
e_{i}^{\phantom{i}\rho }\mathbf{G}f_{T}+\frac{1}{2}e_{i}^{\phantom{i}\rho }%
\left[ \left( f-Tf_{T}\right) \right] +2e_{i}^{\phantom{i}\nu }S_{\nu
}{}^{\mu \rho }\partial _{\mu }({T})f_{TT}=8\pi Ge_{i}^{\phantom{i}\nu }%
\mathcal{T}_{\nu }{}^{\rho }.  \label{ft.31}
\end{equation}

In the case of vacuum, i.e. $\mathcal{T}_{\rho }{}^{\nu }=0$, a vacuum
solution of GR in which $R=0$, also turns out to be a solution of TEGR and
thus of $f\left( T\right) $ gravity without cosmological constant, given
that there exists a frame such the following conditions are satisfied \cite%
{ftSing} 
\begin{equation}
T=0~,~f\left( T\right) _{|T\rightarrow 0}=0  \label{ft.31aa}
\end{equation}%
and%
\begin{equation}
S_{\rho }{}^{\mu \nu }\partial _{\mu }({T})f_{TT}=0.  \label{ft.31bb}
\end{equation}%
\medskip

In a similar way, conditions in the presence of matter can be reconstructed.
Indeed, if $f\left( T\right) _{|T\rightarrow \Lambda }=0$ and $T=-\Lambda $
then the case of GR with cosmological constant is recovered \cite{rfa}.
However, in the presence of a matter source different to that of the
cosmological constant, the additional condition follows 
\begin{equation}
f_{T}\left( T\right) _{|T\rightarrow 0}\neq 0,  \label{ft.31cc}
\end{equation}%
which means that $T\rightarrow 0$ is not a critical point for the function $%
f\left( T\right) .$

This analysis was applied to the case of Bianchi I models in ref.\cite%
{ftSing}. It was found that there exists a family of solutions of the $f(T)$
arbitrary function with a Kasner-like profile in the vacuum setting. The
conclusion being that the Kasner-like model is admitted given that certain
conditions are observed by the free parameters.

In the next section we continue with the determination of the gravitational
field equations in $f\left(T\right)$--gravity for the Bianchi I. These give
the governing equations by which the system can be determined. We study the
stability of the Kasner universe for two functional forms of $%
f\left(T\right) $.

\section{Stability of the Kasner Universe}

\label{stability} \noindent Consider the diagonal nonholonomic frame 
\begin{equation}  \label{ft.b01}
e_{\phantom{i}\mu}^{i}(t)=diag(1,a(t),b(t),c(t)),
\end{equation}
where the corresponding spacetime $g_{\mu\nu}=\eta_{ij}e^{i}_{\phantom{i}%
\mu} e^{j}_{\phantom{j}\nu}$ is that of the Bianchi I spacetime. The
functions $a(t),b(t)$ and $c(t)~$correspond to the three scale factors In
terms of the line element, this takes the following form 
\begin{equation}  \label{ft.b02}
ds^{2}=-dt^{2}+a^{2}(t)dx^{2}+b^{2}\left( t\right) dy^{2}+c\left(
t\right)dz^{2}.
\end{equation}%
\medskip

For the frame in Eq.(\ref{ft.b01}) the invariant $T$ of the Weitzenb\"{o}ck
connection is calculated to be 
\begin{equation}
T=-\frac{2}{abc}\left( c\dot{a}\dot{b}+b\dot{a}\dot{c}+a\dot{b}\dot{c}%
\right) ,  \label{ft.b03}
\end{equation}%
from which we can see that the isotropic scenario of the\ spatially flat
Friedmann--Lema\^{\i}tre--Robertson--Walker spacetime is recovered when, $%
a\left( t\right) =b\left( t\right) =c\left( t\right) ~$\cite{ftSing}.

Hence, in the case of vacuum the gravitational field equations in Eq.(\ref%
{ft.06}) are calculated to be 
\begin{equation}
4f_{T}\left( H_{a}H_{b}+H_{a}H_{c}+H_{b}H_{c}\right) +f=0,  \label{ku.30}
\end{equation}%
\begin{equation}
\dot{T}f_{TT}\left( H_{b}+H_{c}\right) +\frac{f}{2}+f_{T}\left[ \left(
H_{b}+H_{c}\right) \left( H_{a}+H_{b}+H_{c}\right) +\dot{H}_{b}+\dot{H}_{c}%
\right] =0,  \label{ku.31}
\end{equation}%
\begin{equation}
\dot{T}f_{TT}\left( H_{a}+H_{c}\right) +\frac{f}{2}+f_{T}\left[ \left(
H_{a}+H_{c}\right) \left( H_{a}+H_{b}+H_{c}\right) +\dot{H}_{a}+\dot{H}_{c}%
\right] =0,  \label{ku.32}
\end{equation}%
\begin{equation}
\dot{T}f_{TT}\left( H_{a}+H_{b}\right) +\frac{f}{2}+f_{T}\left[ \left(
H_{a}+H_{b}\right) \left( H_{a}+H_{b}+H_{c}\right) +\dot{H}_{a}+\dot{H}_{b}%
\right] =0,  \label{ku.33}
\end{equation}%
where $H_{a}=\frac{\dot{a}\left( t\right) }{a\left( t\right) },~H_{b}=\frac{%
\dot{b}\left( t\right) }{b\left( t\right) }$ and $H_{c}=\frac{\dot{c}\left(
t\right) }{c\left( t\right) }$~are the anisotropic Hubble parameters. The
torsion scalar (or TEGR Lagrangian density) in Eq.(\ref{ft.b03}) is written
equivalently as 
\begin{equation}
T=-2\left( H_{a}H_{b}+H_{a}H_{c}+H_{b}H_{c}\right) ,  \label{ku.34}
\end{equation}%
which also tends to the isotropic value $T=-6H^{2}$ as the scale factors
tend to a single value.

According to the previous section, for a function in which ~$T\rightarrow 0$%
,~and~ $f_{T}\left( T\right) _{T\rightarrow 0}\neq 0$, a power law solution%
\footnote{%
For the conditions of the parameters in a $f\left( T\right) $ theory in
which $f_{T}\left( T\right) _{T\rightarrow 0}=0$ we refer the reader to Ref.%
\cite{ftSing}.}. 
\begin{equation}
a\left( t\right) =a_{0}t^{p_{1}}~,~b\left( t\right)
=b_{0}t^{p_{2}}~,~c\left( t\right) =c_{0}t^{p_{3}},  \label{ku.01}
\end{equation}%
with $a_{0},b_{0},c_{0}$ arbitrary constants, satisfies the gravitational
field equations (\ref{ku.31})-(\ref{ku.33}) and the constraint (\ref{ku.30})
if~Kasner relations holds, i.e. 
\begin{equation}
p_{1}+p_{2}+p_{3}=1\text{ and }\left( p_{1}\right) ^{2}+\left( p_{2}\right)
^{2}+\left( p_{3}\right) ^{2}=1.  \label{ku.02}
\end{equation}%
These relations guarantee that the Kasner metric remains a solution in the
generalized $f(T)$ gravity.

The present study will center investigating the stability of two prominent
functional forms of the $f(T)$ lagrangian within the Kasner-like context.
The Kasner solution has a number of interesting properties that can
elucidate the exotic behavior of a theory in much better way \cite{kasner1}.
The particular functions to be considered are 
\begin{equation}
f_{\left( n\right) }^{A}\left( T\right) =T+f_{0}T^{n},~f_{0}\neq 0,
\label{ku.00}
\end{equation}%
proposed in \cite{Ferraro}~which can take on the role of dark energy in
certain settings, and the exponential theory 
\begin{equation}
f_{\left( n\right) }^{B}\left( T\right) =T+f_{0}\left( 1-e^{-pT^{n}}\right) .
\label{ku.19}
\end{equation}%
The exponential instance was proposed by ref.\cite{Lin2010} for the $n=\frac{%
1}{2}$ subcase, while the $n=1$ was proposed by ref.\cite{bamex}. In both
cases the dark energy analogues were explored.

\subsection{Power-law theory}

For the power--law theory in Eq.(\ref{ku.00}), we consider the condition
that $n\geq 2$ in order to not overlap with the TEGR solution. We split this
up into two analyses with $n=2$ and then $n>2$.

Consider the Kasner solution $\mathbf{a}_{0}\left( t\right) $ where, without
loss of generality in the following, we assume $a_{0}=b_{0}=c_{0}=1$. This
means that the anisotropy will be sourced by the indices of the tetrad
fields. In order to study the stability of this solution we replace the
scale factors with $\mathbf{a}\left( t\right) =\mathbf{a}_{0}\left( t\right)
\left( 1+\varepsilon \mathbf{a}_{\varepsilon }\left( t\right) \right) ,~$%
where $\varepsilon $ is an infinitesimal parameter such that $\varepsilon
^{2}\rightarrow 0$. We then linearize the field equations as $\varepsilon
\rightarrow 0$. In particular with that approach we study the stability of
the trajectories which solves the field equations and describes the Kasner
solution.\textbf{\ }

In order to study the stability of the Kasner solution, in the following we
continue without assuming the constraint condition that $T\left( t\right) $\
vanishes in the perturbations. The perturbation on the scale factors passes
through (\ref{ft.b03}) in $T\left( t\right) $ which reads $T\left( t\right)
=T_{0}\left( t\right) +\varepsilon T_{\varepsilon }\left( t\right) ,~$where $%
T_{0}\left( t\right) $\ denotes the ground state solution which in our case
is zero. 

What do we do, we take a perturbation in the scale factors $a\left( t\right)
=a_{0}\left( t\right) \left( 1+\varepsilon \mathbf{a}_{\varepsilon }\left(
t\right) \right) $; that is, from (\ref{ft.b03}), $\ $\thinspace $T\left(
t\right) $\ reads, $T\left( t\right) =T_{0}\left( t\right) +\varepsilon
T_{\varepsilon }\left( t\right) ,$\ where $T_{0}\left( t\right) $\ is zero;
however we do not impose that $T_{\varepsilon }\left( t\right) $vanishes.
From the perturbation analysis if $a_{\varepsilon }\left( t\right) $\ are
vanishing then necessary $T_{\varepsilon }\left( t\right) $\ reaches zero,
which means that$~$Kasner universe is stable, otherwise when $T_{\varepsilon
}\left( t\right) \neq 0$, the ground state solution (\ref{ku.01}), (\ref%
{ku.02}) is unstable.

\subsubsection{Case $n=2$}

The lagrangian density takes on the form $f_{\left( 2\right) }\left(
T\right) =T+f_{0}T^{2}$ in this setting. Thus, taking the perturbation
described above in the field equations, results in the following
non-autonomous dynamical system: 
\begin{align}
0& =4t^{3}\ddot{a}_{\varepsilon }-\left( 32f_{0}\left( p_{1}-1\right)
p_{1}-2\left( 1+3p_{1}\right) t^{2}\right) \dot{a}_{\varepsilon }  \notag
\label{ku.03} \\
& -\left( -16f_{0}p_{1}\left( 1+p_{1}+\Delta \left( p_{1}\right) \right)
+\left( 1-3p_{1}+\Delta \left( p_{1}\right) \right) t^{2}\right) \dot{b}%
_{\varepsilon }+  \notag \\
& -\left( -16f_{0}p_{1}\left( 1+p_{1}-\Delta \left( p_{1}\right) \right)
+\left( 1-3p_{1}-\Delta \left( p_{1}\right) \right) t^{2}\right) \dot{c}%
_{\varepsilon },
\end{align}%
\begin{align}
0& =4t^{3}\ddot{b}_{\varepsilon }\left( t\right) -\left( 2\left(
8f_{0}\left( p_{1}-1\right) \left( 1-p_{1}-\Delta \left( p_{1}\right)
\right) \right) +\Delta \left( p_{1}\right) t^{2}\right) \dot{a}%
_{\varepsilon }+  \notag  \label{ku.04} \\
& -\left( 16f_{0}p_{1}\left( 1-p_{1}+\Delta \left( p_{1}\right) \right)
-\left( 5-3p_{1}-3\Delta \left( p_{1}\right) \right) t^{2}\right) \dot{b}%
_{\varepsilon }+  \notag \\
& -\left( 16f_{0}\left( p_{1}\left( 2p_{1}-1\right) -1+\Delta \left(
p_{1}\right) \right) -\left( 1-3p_{1}-\Delta \left( p_{1}\right) \right)
t^{2}\right) \dot{c}_{\varepsilon },
\end{align}%
\begin{align}
0& =4t^{3}\ddot{c}_{\varepsilon }\left( t\right) +2a^{\prime }(t)\left(
\Delta \left( p_{1}\right) t^{2}-8f_{0}(p_{1}-1)\left( 1-p1+\Delta \left(
p_{1}\right) \right) \right) \dot{a}_{\varepsilon }+  \notag  \label{ku.05}
\\
& +b^{\prime }(t)\left( 16f_{0}\left( p_{1}\left( 1-2p_{1}\right) +\Delta
\left( p_{1}\right) +1\right) +\left( 1-3p_{1}+\Delta \left( p_{1}\right)
\right) t^{2}\right) \dot{b}_{\varepsilon }+  \notag \\
& +\left( 16f_{0}p_{1}\left( p_{1}-1+\Delta \left( p_{1}\right) \right)
+\left( 5-3p_{1}+3\Delta \left( p_{1}\right) \right) t^{2}\right) \dot{c}%
_{\varepsilon },
\end{align}%
where the constraint equation takes on the form 
\begin{equation}
2\left( 1-p_{1}\right) \dot{a}_{\varepsilon }\left( t\right) +\left(
1+p_{1}+\Delta \left( p_{1}\right) \right) \dot{b}_{\varepsilon }+\left(
1+p_{1}-\Delta \left( p_{1}\right) \right) \dot{c}_{\varepsilon }=0,
\label{ku.06}
\end{equation}%
with 
\begin{equation*}
\Delta \left( p_{1}\right) =\sqrt{1+\left( 2-3p_{1}\right) p_{1}}.
\end{equation*}%
\medskip

In the above system, for convenience, we consider the change of variables 
\begin{equation}
a_{\varepsilon }\left( t\right) \rightarrow \int \alpha \left( t\right)
dt~,~b_{\varepsilon }\left( t\right) \rightarrow \int \beta \left( t\right)
dt~\text{,~}c\left( t\right) \rightarrow \int \gamma \left( t\right) dt,
\label{ku.07}
\end{equation}%
since the system reduces to a first-order algebraic-differential system
whose analytical solution can easily be determined. For the lagrangian under
consideration the particular values $p_{1}=1$ and $p_{1}=-\frac{1}{3}$ are
assumed so that the critical cases of the Bianchi I spacetime can represent
the Minkowski spacetime in nonstandard coordinates or a locally rotational
spacetime (LRS), respectively.

For $p_{1}=1$ it follows that $p_{2}=p_{3}=0$; then, the analytic solution
of the linearized system is given by 
\begin{equation}
\alpha \left( t\right) =\alpha _{0}t^{-2}~,~~\beta \left( t\right) =\beta
_{0}t^{-1}~,~\gamma \left( t\right) =-\beta _{0}t^{-1},  \label{ku.08}
\end{equation}%
that is\footnote{%
Without loss of generality, we omit the integration constants.} 
\begin{equation}
a_{\varepsilon }\left( t\right) \simeq t^{-1}~,~b_{\varepsilon }\left(
t\right) \simeq \ln t~,~c_{\varepsilon }\left( t\right) \simeq \ln t.
\label{ku.09}
\end{equation}%
Hence, the perturbations admit a logarithmic singularity, that is, one which
results in an unstable Kasner universe.

Secondly, for the $p_{1}=-\frac{1}{3}$ case we find that the exact solution
of the linearized system to be 
\begin{equation}
\alpha \left( t\right) =\alpha _{0}t^{-2}~,~~\beta \left( t\right) =-2\alpha
_{0}t^{-2}+\beta _{0}t^{-1}~,~\gamma \left( t\right) =-2\alpha
_{0}t^{-2}-\beta _{0}t^{-1},  \label{ku.10}
\end{equation}%
which means that the perturbation equations evolve as 
\begin{equation}
a_{\varepsilon }\left( t\right) =-a_{0}t^{-1}~,~b_{\varepsilon }\left(
t\right) \simeq 2a_{0}t^{-1}+\beta _{0}\ln t~,~c_{\varepsilon }\left(
t\right) \simeq 2a_{0}t^{-1}-\beta _{0}\ln t.  \label{ku.11}
\end{equation}%
From Eq.(\ref{ku.11}), we observe that for a set of initial conditions, the
Kasner solution can be configured to be stable. However, in general the
solution is unstable and the point which describes the Kasner universe is a
saddle point.

Finally, for other values of $p_{1}$ the following perturbation equations
are found 
\begin{equation}
a_{\varepsilon }\left( t\right) \simeq \bar{a}_{0}t^{-1}+\bar{a}_{1}\ln
t~,~b_{\varepsilon }\left( t\right) \simeq \bar{b}_{0}t^{-1}+\bar{b}_{1}\ln
t~,~c_{\varepsilon }\left( t\right) \simeq \bar{c}_{0}t^{-1}+\bar{c}_{1}\ln
t.  \label{ku.12}
\end{equation}%
where there are only two free integration constants. That means that $\bar{a}%
_{i},\bar{b}_{i}$ and $\bar{c}_{i}$ are functions of two integration
constants $I_{1}$,~$I_{2}$, and of the parameter $p_{1}$. If $\bar{a}_{1}=0$
then it follows that~$\bar{b}_{1}=\bar{c}_{1}=0$. We conclude that the
Kasner solution is unstable and the point that describes the Kasner universe
is a saddle point.

\subsubsection{Case $n>2$}

The situation changes in an important way for the $n>2$ case since $%
f_{,TT}\left( T\right) _{T\rightarrow 0}=0$, which means that $T=0$ is an
inflection point for the lagrangian. This is the main distinction between
the two ranges of $n$. The perturbation equations turn out to be 
\begin{align}
0& =2t\left( \ddot{b}_{\varepsilon }(t)+\ddot{c}_{\varepsilon }(t)\right)
+\left( 3\left( 1-p_{1}\right) -\Delta \left( p_{1}\right) \right) \dot{b}%
_{\varepsilon }(t)+  \notag  \label{ku.13} \\
& ~~~+\left( 3\left( 1-p_{1}\right) +\Delta \left( p_{1}\right) \right)
c_{\varepsilon }(t),
\end{align}%
\begin{align}
0& =2t\left( \ddot{a}_{\varepsilon }(t)+\ddot{c}_{\varepsilon }(t)\right)
+2\left( 1+\Delta \left( p_{1}\right) \right) \dot{c}_{\varepsilon }(t)+ 
\notag  \label{ku.14} \\
& +\left( 1+3p_{1}+\Delta \left( p_{1}\right) \right) \dot{a}_{\varepsilon
}(t),
\end{align}%
\begin{align}
0& =2t\left( \ddot{a}_{\varepsilon }(t)+\ddot{b}_{\varepsilon }(t)\right)
+2\left( 1-\Delta \left( p_{1}\right) \right) \dot{b}_{\varepsilon }(t)+ 
\notag  \label{ku.15} \\
& +\left( 1+3p_{1}-\Delta \left( p_{1}\right) \right) \dot{a}_{\varepsilon
}(t),
\end{align}%
while the constraint equation is given by, 
\begin{equation}
0=2(1-p_{1})\dot{a}_{\varepsilon }(t)+\left( 1+p_{1}+\Delta \left(
p_{1}\right) \right) \dot{b}_{\varepsilon }(t)+\left( 1+p_{1}-\Delta \left(
p_{1}\right) \right) \dot{c}_{\varepsilon }(t).  \label{ku.16}
\end{equation}%
\medskip

We observe that the system of equations in Eq.(\ref{ku.13})-(\ref{ku.16})
differs from the $n=2$ case, with the most important difference being that
the second set of relations are independent of the index $n$. Moreover, the
transformations in Eq.(\ref{ku.07}) can also be applied here to reduce the
system to a first-order, algebraic differential system with a
straightforward solution.

To solve the present system we adopt a different strategy where the cosmic
time coordinate is transformed through $t\rightarrow \tau $, so that $%
t=e^{\tau }$. Consequently, we find that $\frac{da}{dt}=e^{-\tau }\frac{da}{%
d\tau },$ and $\frac{d^{2}a}{dt^{2}}=e^{-2\tau }\left( \frac{d^{2}a}{d\tau
^{2}}-\frac{da}{d\tau }\right) $. The change of variables changes the
dynamical system in Eq.(\ref{ku.13})-(\ref{ku.16}) to an autonomous system
which means that the critical points can be analyzed.

To do this, we define a new set of variables $x_{1}\left( \tau \right)
,~x_{2}\left( \tau \right) ~$and $x_{3}\left( \tau \right) $ with 
\begin{equation}
a_{\varepsilon }=\frac{1}{2}\left( x_{1}+x_{2}-x_{3}\right)
~,~b_{\varepsilon }\left( \tau \right) =\frac{1}{2}\left(
x_{1}-x_{2}+x_{3}\right) ~,~c_{\varepsilon }\left( \tau \right) =\frac{1}{2}%
\left( -x_{1}+x_{2}+x_{3}\right) ,  \label{ku.17}
\end{equation}%
and $x_{3+i}=\frac{dx_{i}}{d\tau }$. Therefore the system in Eq.(\ref{ku.13}%
)--(\ref{ku.15}) can be written in the form 
\begin{equation}
\frac{d\mathbf{x}}{d\tau }\mathbf{=Ax},  \label{ku.18}
\end{equation}%
where $\mathbf{A}$ is a $6\times 6$ matrix which has a positive Eigenvalue.
This means that the critical point $\mathbf{x=0}$ describing the Kasner
solution is unstable. Specifically we find that this is a saddle point.
Additionally, we find that the eigenvalues of $\mathbf{A}$ are independent
on $p_{1}$ which means that the solution is unstable for the LRS spacetime
as well.

\subsection{Exponential theory}

As discussed in section \ref{field}, in order to recover the TEGR solution
in the exponential theory \ref{ku.19}, the power $n$ must satisfy $n\geq 1$,
which follows from Eq.\ref{ft.31bb}. This means that the model given in ref.%
\cite{Lin2010} does not recover the limit of GR, which contradicts the
picture presented in ref.\cite{bamex}. As in the previous case, we perform
our analysis separately for $n=1$ and for $n>1$.

\subsubsection{Case $n>1$}

Considering the $n>1$ instance for the exponential theory scenario, we
perform the perturbation analysis again with the Kasner solution. The same
perturbation equations result as in Eq.(\ref{ku.13})--(\ref{ku.16}) which
means that the previous analysis holds, and that the Kasner solution is
again unstable here.

However, that coincidence it is not a surprise. Indeed a series expansion of
(\ref{ku.19}) around the Kasner solution where $T=0$, gives the expansion 
\begin{equation}
f\left( T\right) =\left( 1-f_{0}p\right) T+\varepsilon f_{0}pT^{n}+O\left(
\varepsilon ^{2}\right) ,  \label{ku.20}
\end{equation}%
which is in the form of the power law theory (\ref{ku.00}), and results in
the previous setup.\medskip

We proceed with the $n=1$ case which as we will see has similarities with
the quadratic power law theory.

\subsubsection{Case $n=1$}

As in the quadratic case, $f_{T}\left( T\right) _{|T\rightarrow 0}\neq 0$
follows while $f_{0}\neq -p^{-1}$. Thus, for the exponential lagrangian
case, Eq.(\ref{ku.19}), with $n=1$, the linearized equations around the
Kasner solution are given from the following linear non-autonomous system of
second-order differential equations

\begin{align}  \label{ku.22}
0 & =-4(1+f_{0}p)t^{3}\ddot{a}_{\varepsilon}\left(t\right)+2a^{\prime}(t)%
\left(8f_{0}p^{2}(p_{1}-1)p_{1}-(1+f_{0}p)(3p_{1}+1)t^{2}\right)+  \notag \\
&+\dot{b}_{\varepsilon}(t)\left(-f_{0}p\left(
8pp_{1}\left(p_{1}+\Delta\left( p_{1}\right)+1\right)
+\left(1-3p_{1}+\Delta\left(p_{1}\right) \right)t^{2}\right) -\left(
1-3p_{1}+\Delta\left(p_{1}\right) \right) t^{2}\right)+  \notag \\
&+\dot{c}_{\varepsilon}(t)\left(-f_{0}p\left(8pp_{1}\left(1+p_{1}-\Delta%
\left( p_{1}\right) \right) +\left(1-3p_{1}-\Delta\left(p_{1}\right) \right)
t^{2}\right)-\left(1-3p_{1}-\Delta\left(p_{1}\right) \right)t^{2}\right),
\end{align}
\begin{align}  \label{ku.23}
0 & =-4(1+f_{0}p)t^{3}\ddot{b}_{\varepsilon}\left(t\right) +2\dot{a}%
_{\varepsilon}(t)\left(-f_{0}p\left(4pp_{1}-1)\left(p_{1}+\Delta\left(p_{1}%
\right)-1\right) +\Delta\left( p_{1}\right) t^{2}\right)-\Delta\left(
p_{1}\right) t^{2}\right)+  \notag \\
&+\dot{b}_{\varepsilon}(t)\left(-f_{0}p\left(-8pp_{1}\left(1+p_{1}-\Delta%
\left(p_{1}\right) \right)-\left(5-3p_{1}-3\Delta\left(p_{1}\right)\right)
t^{2}\right) +\left(5-3p_{1}-3\Delta\left(p_{1}\right)\right)t^{2}\right) + 
\notag \\
&+\dot{c}_{\varepsilon}(t)\left(-f_{0}p\left(
8p\left(1+p_{1}\left(1-2p_{1}\right)-\Delta\left( p_{1}\right)
\right)-\left(1-3p_{1}-\Delta\left(p_{1}\right)\right)t^{2}\right) +\left(
1-3p_{1}-\Delta\left(p_{1}\right)\right)t^{2}\right),
\end{align}
\begin{align}  \label{ku.24}
0 & =-4(1+f_{0}p)t^{3}\ddot{c}_{\varepsilon}\left(t\right)+2\dot{a}%
_{\varepsilon}(t)\left(-f_{0}p\left(-4p(p_{1}-1)\left(1-p_{1}+\Delta\left(
p_{1}\right)\right)-\Delta\left(p_{1}\right) t^{2}\right)+\Delta\left(
p_{1}\right) t^{2}\right)+  \notag \\
&+\dot{b}_{\varepsilon}(t)\left(-f_{0}p\left(8p\left(1+p_{1}\left(1-2p_{1}%
\right)+\Delta\left(p_{1}\right)\right)-\left(1-3p_{1}+\Delta\left(p_{1}%
\right) \right) t^{2}\right)+\left(1-p_{1}+\Delta\left(p_{1}\right) \right)
t^{2}\right)+  \notag \\
&+\dot{c}_{\varepsilon}(t)\left(
-f_{0}p\left(8pp_{1}\left(p_{1}-1+\Delta\left(p_{1}\right)\right)
-\left(5-3p_{1}+3\Delta\left(p_{1}\right)\right)t^{2}\right)+%
\left(5-3p_{1}+3\Delta\left(p_{1}\right) \right)t^{2}\right),
\end{align}
where the constraint equation is found to be 
\begin{equation}  \label{ku.21}
0=(1+f_{0}p)\left(2(p_{1}-1)\dot{a}_{\varepsilon}(t)-\left(1+p_{1}+\Delta%
\left(p_{1}\right)\right)\dot{b}_{\varepsilon}(t)-\left(1+p_{1}-\Delta%
\left(p_{1}\right)\right)\dot{c}_{\varepsilon}(t)\right).
\end{equation}%
\medskip

Applying the constraint equations reduces the system Eq.(\ref{ku.22})--(\ref%
{ku.24}) to the following pair of first--order differential equations 
\begin{equation}
-\left( 1-p_{1}+\Delta \left( p_{1}\right) \right) \frac{d\alpha \left( \tau
\right) }{d\tau }+p_{1}\left( 1-3p_{1}+\Delta \left( p_{1}\right) \right)
\alpha \left( \tau \right) +2p_{1}\Delta \left( p_{1}\right) \beta (\tau )=0,
\label{ku.25}
\end{equation}%
\qquad 
\begin{equation}
\left( 1+p_{1}-\Delta \left( p_{1}\right) \right) \frac{d\beta \left( \tau
\right) }{d\tau }+p_{1}\left( 3\left( 1-p_{1}\right) -\Delta \left(
p_{1}\right) \right) \alpha \left( \tau \right) +(1-p_{1})\left(
1+3p_{1}-\Delta \left( p_{1}\right) \right) \beta (\tau )=0,  \label{ku.26}
\end{equation}%
where we have performed the change of variables~$\alpha \left( t\right) =%
\dot{a}_{\varepsilon }\left( t\right) ,~\beta \left( t\right) =\dot{\beta}%
_{\varepsilon }\left( t\right) $ and $t=e^{\tau }$.

The dynamical system in Eq.(\ref{ku.25})--(\ref{ku.26}) admits the following
family of critical points 
\begin{equation}
\beta \left( \tau \right) =\frac{1-3p_{1}+\Delta \left( p_{1}\right) }{%
2\Delta \left( p_{1}\right) }\alpha \left( \tau \right) ,  \label{ku.27}
\end{equation}%
while the eigenvalues of the matrix which defines the system Eq.(\ref{ku.25}%
)-(\ref{ku.26}) are $e_{1}=0$ and $e_{2}=-1$. Since the dynamical system is
linear, an analytic solution can easily be found in the form 
\begin{equation}
\alpha \left( \tau \right) =\alpha _{0}+\alpha _{1}e^{-\tau }~,~b\left( \tau
\right) =\beta _{0}+\beta _{1}e^{-\tau },  \label{ku.28}
\end{equation}%
where $\alpha _{1}=\alpha _{1}\left( \alpha _{0},\beta _{0},p_{1}\right) $
and $\beta _{1}=\beta _{1}\left( \alpha _{0},\beta _{0},p_{1}\right) $. It
therefore follows that the perturbations should take the form 
\begin{equation}
a_{\varepsilon }\left( t\right) =\alpha _{0}t+a_{1}\ln t~,~b_{\varepsilon
}\left( t\right) =\beta _{0}t+\beta _{1}\ln t~,~c_{\varepsilon }\left(
t\right) =c_{0}t+c_{1}\ln t~.  \label{ku.29}
\end{equation}%
Hence, for $n=1$ the point which describes the Kasner solution is a
hyperbolic point. Hence, the Kasner universe is not stable for any initial
condition unlike for the power-law situation in which the Kasner solution is
described by a saddle point.

\section{De Sitter Universe}

\label{deSitter} In the previous section we show that the Kasner universe it
is not a stable solution of the $f(T)$ field equations. Here, we will
demonstrate that the de Sitter Universe is an attractor for the
theory.\medskip\ The conditions for the existence and stability of the de
Sitter metric in modified gravity theories is an important factor in the
evaluation of their ability to explain the observed late-time acceleration
of the universe or accommodate an early period of inflation that creates
isotropic expansion from a wide range of anisotropic and inhomogeneous prior
conditions.

The gravitational field equations Eq.(\ref{ku.30})--(\ref{ku.33}) offer an
algebraic-differential system of first-order differential equations with
respect to the Hubble rate variables $H_{a},~H_{b}$ and $H_{c}$. We are
interested in the critical points of this system. To do this, consider the
power-law lagrangian density in Eq.(\ref{ku.00}). Note that any critical
point of the system Eq.(\ref{ku.30})--(\ref{ku.33}) describes constant
expansion parameters $H_{a},~H_{b}$ and $H_{c}$; that is, exponential scale
factors 
\begin{equation}
a\left( t\right) =a_{0}e^{H_{a}^{P}t}~,~b\left( t\right)
=b_{0}e^{H_{b}^{P}t}~,~c\left( t\right) =c_{0}e^{H_{c}^{P}t}.  \label{ku.0aa}
\end{equation}%
\medskip

\begin{figure}[tbp]
\includegraphics[height=8cm]{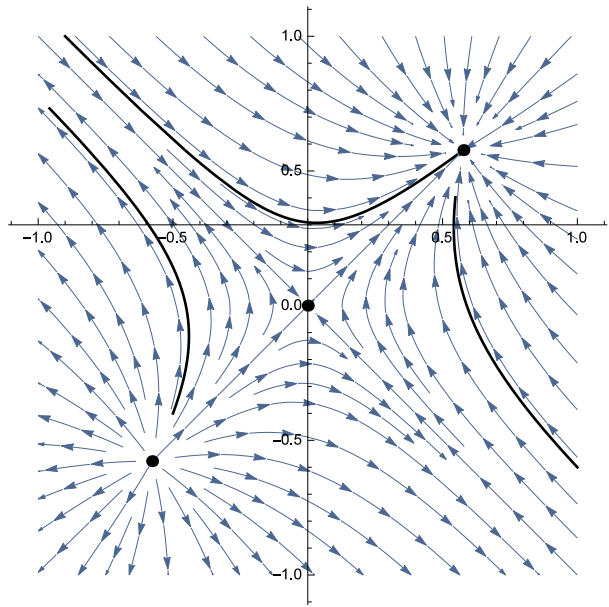}\centering
\caption{Phase portrait in the plane $H_{a}-H_{b}$ for the quadratic $%
f\left( T\right) =T+f_{0}T^{2}$ lagrangian, with $f_{0}=\frac{1}{12}$. Solid
lines describe real solutions of the field equations.}
\label{plot1}
\end{figure}

For $n=2$, i.e. the quadratic case, the dynamical system admits the
following (real) critical points: 
\begin{equation}
P_{\pm }:\left( H_{a},H_{b},H_{c}\right) =\left( _{\pm }\frac{1}{6\sqrt{f_{0}%
}},_{\pm }\frac{1}{6\sqrt{f_{0}}},_{\pm }\frac{1}{6\sqrt{f_{0}}}\right) ,
\end{equation}%
and 
\begin{equation}
P_{0}:\left( H_{a},H_{b},H_{c}\right) =\left( 0,0,0\right) .
\end{equation}%
All these points describe isotropic universes: $P_{0}$ describes the
Minkowski spacetime and $P_{\pm }$ describes de Sitter solutions. At the
point $P_{0}$, the geometric dark energy fluid vanishes, while at the other
points it mimics the cosmological constant. It is straightforward to see
that the point $P_{-}$ is always unstable, while $P_{+}$ is a future
attractor. Finally, the point $P_{0}~$has three zero eigenvalues. From Fig.(%
\ref{plot1}), we can see that $P_{0}$ is a saddle point, and indeed the
unique attractor is the $P_{+}$ point.\medskip

It is well known that in the presence of the cosmological constant, the
Bianchi I universe becomes asymptotically isotropic at late times and, in
the cosmological scenario, $f\left( T\right) $ drives the isotropisation of
the anisotropic spacetime.\medskip

Finally, for higher-order powers of the index $n$, we find that there exists
a de Sitter attractor if and only if, 
\begin{equation}
sign\left( f_{0}\right) =\left( -1\right) ^{n}.
\end{equation}

\section{Discussion}

\label{conc} The main subject of this study was to prove the existence, and
study the stability, of the Kasner vacuum solution of Bianchi type I for the
modified $f\left( T\right) $ teleparallel gravity. Vacuum solutions have not
been studied extensively in \ $f\left( T\right) $ gravity, and it was only
recently found that there exists an anisotropic vacuum Bianchi I exact
solution \cite{ftSing} in this gravitational context.

In particular, we found conditions under which a nonlinear $f\left( T\right) 
$ gravity provides solutions of TEGR and discussed their features. Those
conditions applied in two functional forms for $f\left( T\right) $ gravity
where there was a special interest in the dark energy section; specifically,
for the power-law $f(T)$ model (\ref{ku.00}) and the exponential $f(T)$
model (\ref{ku.19}). The assumption that the Kasner universe solves the
field equations was found to place a number of constraints on the free
parameters of the $f\left( T\right) $~models in question. The power-law
model (\ref{ku.00}) admits the Kasner solution only if the power in $T^{n}$
satisfies $n\geq 2$; while for the exponential model (\ref{ku.19}), it was
found that the exponential model presented in ref.\cite{Lin2010} does not
yield a Kasner universe solution.

We performed a perturbation of the Kasner solution and study the linearized
field equations in a number of cases. We solved the perturbation equations
directly and analysed for the critical points. We found that the
perturbation equations admit a logarithmic singularity, which correlates
with the fact that the fixed-point analysis shows the Kasner solution to be
a saddle point. Therefore, the Kasner universe turns out to be unstable in $%
f\left( T\right) $ gravity.

\begin{figure}[ptb]
\includegraphics[height=5cm]{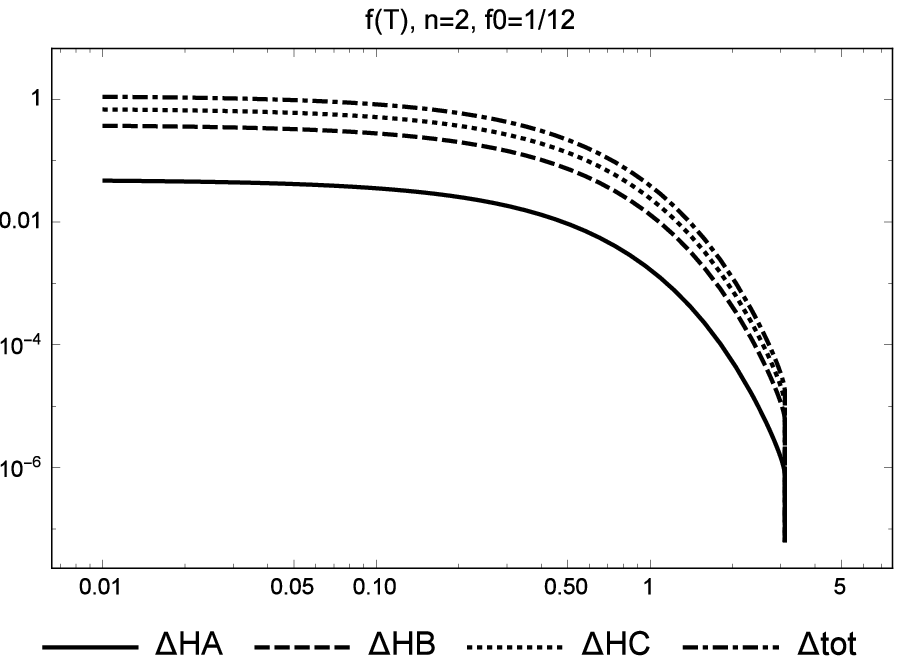}\centering
\includegraphics[height=5cm]{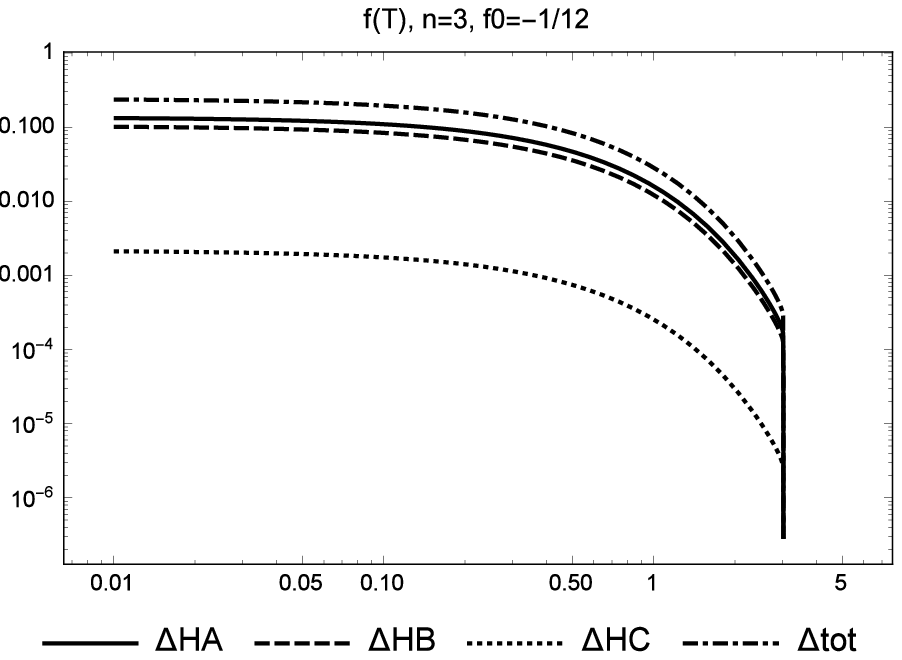}\centering
\caption{Numerical simulation of the anisotropic Hubble parameters$~\Delta H$
for the power-law model, plotted against time (\protect\ref{ku.00}). Figures
are for $n=2\,\ $(left) and $n=3~$(right) for initial conditions close to
the Kasner solution. We observe that the de Sitter Universe is a late-time
attractor.}
\label{ppl1}
\end{figure}

Additionally, in order to better understand the evolution of the field
equations, a fixed-point analysis was performed which resulted in the
power-law model providing a future attractor which was an isotropic de
Sitter universe. Therefore, we can claim that in $f\left( T\right) $
gravity, small fluctuations in the (anisotropic) Kasner universe can lead
evolve towards an accelerating de Sitter universe. This is demonstrated in
the phase portrait shown in Fig. \ref{ppl1}, where the anisotropic Hubble
parameters 
\begin{equation}
\Delta H_{a}=\frac{\left( H_{a}-H_{V}\right) ^{2}}{\left( H_{V}\right) ^{2}}%
,~\Delta H_{b}=\frac{\left( H_{b}-H_{V}\right) ^{2}}{\left( H_{V}\right) ^{2}%
}~,~\Delta H_{c}=\frac{\left( H_{c}-H_{V}\right) ^{2}}{\left( H_{V}\right)
^{2}},
\end{equation}%
and 
\begin{equation}
\Delta H_{tot}=\Delta H_{a}+\Delta H_{b}+\Delta H_{c},
\end{equation}%
are presented, where $H_{V}$ is the Hubble parameter for the average volume, 
$H_{V}=\frac{\dot{V}}{V}$, in which the volume factor is $V\left( t\right)
\equiv a\left( t\right) b\left( t\right) c\left( t\right) $.

The numerical simulation shown in Fig.(\ref{ppl1}) is for the power-law
model (\ref{ku.00}) for the cases of $n=2$, $3,$ with initial conditions
close to that of Kasner universe. We observe that the anisotropic Hubble
parameters ultimately vanish to high accuracy which means that the spacetime
evolves to a de Sitter phase, again further adding to the status of the
Kasner universe within the $f(T)$ context.

For completeness, let us consider now the $f\left( T\right) =T^{n}$ theory
for $n\geq 2$. Now, $f_{T}\left( T\right) _{|T\rightarrow 0}=0$, where the
scale factors (\ref{ku.01}) solve the field equations if and only if%
\footnote{%
We consider that at least two of the $p$'s are not equal.} 
\begin{equation}
p_{1}p_{2}+p_{1}p_{3}+p_{2}p_{3}=0.  \label{ks.01}
\end{equation}%
As before we perform a linear perturbation on the power-law solution (\ref%
{ku.01}), where we see that the perturbations depend on the power $n$ of the
theory. In particular, for $n>2$, in which $f_{T}\left( T\right)
_{|T\rightarrow 0}$ and $f_{TT}\left( T\right) _{|T\rightarrow 0}=0$, we
find that the linear perturbations of the field equations are satisfied
identically (unlike in the $n=2$ case), with $f_{TT}\left( T\right)
_{|T\rightarrow 0}=2,~$where the following differential equation follows:%
\begin{eqnarray}
0 &=&\left( p_{1}+p_{2}\right) t\left( \left( p_{1}+p_{2}\right) \dot{h}%
_{c}+p_{2}\dot{h}_{a}+p_{1}\dot{h}_{b}\right) +  \notag \\
&&+\left( p_{1}+p_{2}\right) \left( p_{1}\left( p_{1}-3\right) +p_{2}\left(
p_{2}-3\right) +p_{1}p_{2}\right) h_{c}+  \notag \\
&&+\left( \left( p_{1}\right) ^{2}+p_{1}\left( p_{2}-3\right) +p_{2}\left(
p_{2}-3\right) \right) \left( p_{2}h_{a}+p_{1}h_{b}\right)  \label{ks.02}
\end{eqnarray}%
where $\mathbf{h}\left( t\right) $ are the perturbations of the Hubble
functions for the three different directions; that is, we have considered
the perturbation $\mathbf{H}\left( t\right) =\mathbf{H}_{0}\left( t\right)
\left( 1+\varepsilon \mathbf{h}\left( t\right) \right) ,$ in which $\mathbf{H%
}_{0}\left( t\right) $ denotes the zero-order solution (\ref{ku.01}). \ For $%
n>2,$ we search for higher-order corrections, and we find that for the $%
n^{th}$ correction equation that (\ref{ks.02}) follows.

\begin{figure}[tbp]
\includegraphics[height=7cm]{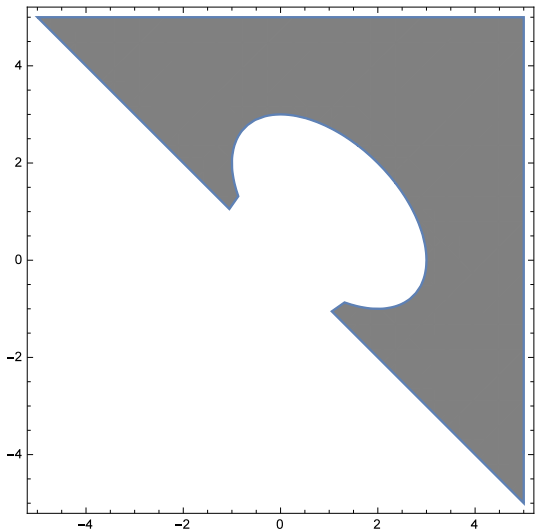}\centering
\caption{Points on the plane $\left( p_{1},p_{2}\right) $ for which the
perturbation (\protect\ref{ks.032}) decays so that the Kasner-like solution (%
\protect\ref{ks.01}) is stable when $h_{a}\left( t\right) ,~h_{b}\left(
t\right) $ decays as a power of $t^{-1}$. }
\label{fig0001}
\end{figure}

From the above, we conclude that for\ the $f\left( T\right) =T^{n},$ with $%
n\geq 2,$ the stability of the Kasner Universe (\ref{ku.02}) and of the
Kasner-like solution (\ref{ks.01}) depend on the nature of the
perturbations. For instance, if we assume that $h_{a}$ and $h_{b}$ decay
with a power of $t^{-1}$, then $h_{c}\left( t\right) $ is given by the
expression%
\begin{equation}
h_{c}\left( t\right) \simeq t^{A\left( p_{1},p_{2}\right)
}+h_{c}^{0}t^{B\left( p_{1,}p_{2}\right) },  \label{ks.032}
\end{equation}%
from which it follows that $A\left( p_{1},p_{2}\right) $ and $B\left(
p_{1},p_{2}\right) $ are negative constants iff $p_{1},p_{2}$ have values
from the grey grid in Fig. \ref{fig0001}

By comparing our results for the $f\left( T\right) $ theory of gravity with
those for the $f\left( R\right) $ theories of gravity \cite{barcl} we see
that, while there are some similarities in the existence of Kasner-like
solutions, there are differences between these two families of theories with
regard to the stability of the Kasner-like solution. Such differences have
been observed before in the case of an isotropic universe, (see the
discussion in \cite{ftSing}), and those differences are expected between
these two different classes of theories because of the different number of
degrees of freedom that they each support.

\begin{acknowledgments}
AP acknowledges financial support \ of FONDECYT grant no. 3160121. JDB is
supported by the Science and Technology Funding Council (STFC) of the United
Kingdom.
\end{acknowledgments}


\begin{thebibliography}{99}
\bibitem{Riess:1998cb} A.~G. Riess et~al., Astron. J. \textbf{116}, 1009
(1998).

\bibitem{Perlmutter:1998np} S. Perlmutter et al., Astrophys. J. \textbf{517}%
, 565 (1999).

\bibitem{clifton} T. Clifton, P.G. Ferreira, A. Padilla and C. Skordis,
Phys. Rept. \textbf{513,} 1 (2012)

\bibitem{weinberg1972gravitation} S. Weinberg, \textit{Gravitation and
Cosmology: Principles and Applications of the General Theory of Relativity}
(Wiley, New York, 1972).

\bibitem{Martin:2012bt} J. Martin, Comptes Rendus Physique \textbf{13}, 566
(2012)

\bibitem{barott} J.D. Barrow and A.C. Ottewill, J. Phys. A. \textbf{16,}
2757 (1983)

\bibitem{ein28} A. Einstein 1928, Sitz. Preuss. Akad. Wiss. p. 217; ibid p.
224

\bibitem{Tsamp} M. Tsamparlis, Phys. Lett. A \textbf{75,} 27 (1979)

\bibitem{Ferraro} G.R. Bengochea and R. Ferraro, Phys. Rev. D \textbf{79,}
124019 (2009)

\bibitem{Lin2010} E.V.~Linder, Phys.\ Rev.\ D \textbf{81}, 127301 (2010)

\bibitem{sar01} M. Kr\v{s}\v{s}\'{a}k and E.N.\ Saridakis, Class.\ Quantum
Grav. \textbf{33,} 115009 (2016)

\bibitem{SLB} B. Li, T.P. Sotiriou and J.D. Barrow, Phys. Rev. D \textbf{83}%
, 064035 (2011)

\bibitem{SLB2} B. Li, T.P. Sotiriou and J.D. Barrow, Phys. Rev. D \textbf{83}%
, 104030 (2011)

\bibitem{kasner1} E. Kasner, Am. J. Math. \textbf{43,} 217 (1921)

\bibitem{DemaretKL} J. Demaret, M. Henneaux, and P. Spindel, Phys. Lett. B 
\textbf{164,} 27 (1985)

\bibitem{cornish} N. Cornish and J. Levin, Phys. Rev. D \textbf{55}, 7489
(1997)

\bibitem{coleyb} A.A. Coley, \textit{Dynamical Systems and Cosmology},
(Springer Science and Business Media, Dordrecht, 2003)

\bibitem{kas1} K. Adhav, A. Nimkar, R. Holey, Int. J. Theor. Phys. \textbf{%
46,} 2396 (2007)

\bibitem{kas2} S.M.M. Rasouli, M. Farhoudi and H.R. Sepangi, Class. Quantum
Grav. \textbf{28,} 155004 (2011)

\bibitem{kas3} X.O. Camanho, N. Dadhich and A. Molina, Class. Quantum Grav. 
\textbf{32,} 175016 (2015)

\bibitem{kas4} P. Halpern, Phys. Rev. D \textbf{63,} 024009 (2001)

\bibitem{kas5} M.V. Battisti and G. Montani, Phys. Lett. B \textbf{681,} 179
(2009)

\bibitem{kas6} K. Andrew, B. Golen and C.A. Middleton, Gen. Relativ. Gravit. 
\textbf{39, }2061 (2007)

\bibitem{kas7} S.A. Pavluchenko, Phys. Rev. D \textbf{94,} 024046 (2016)

\bibitem{barcl} J.D. Barrow and T. Clifton, Class Quantum Grav. \textbf{23,}
L1 (2006)

\bibitem{barcl2} T. Clifton and J.D.\ Barrow, Class Quantum Grav. \textbf{23,%
} 2951 (2006)

\bibitem{topo} A. Toporensky and D. M\"{u}ller, Gen. Relativ. Gravit. 
\textbf{49,} 8 (2017)

\bibitem{bar01} J.D. Barrow and J. Middleton, Phys. Rev. D \textbf{75,}
123515 (2007)

\bibitem{zs} Y.B Zeldovich and A.A. Starobinsky, Sov. Phys. JETP \textbf{34}%
, 1159 (1972)

\bibitem{bt} J.D. Barrow and M.S. Turner, Nature \textbf{291}, 469 (1981)

\bibitem{bmass} J.D. Barrow, Nucl. Phys. B \textbf{208}, 501 (1982)

\bibitem{bher} J.D. Barrow and S. Hervik, Phys. Rev. D \textbf{81}, 023513,
(2010)

\bibitem{silk} J.D. Barrow, P. G. Ferreira and J. Silk, Phys. Rev. Lett. 
\textbf{78}, 3610 (1997)

\bibitem{skew} J.D. Barrow, Phys. Rev. D \textbf{55}, 7451 (1997)

\bibitem{HT} S.W. Hawking and R.J. Tayler, Nature \textbf{209}, 1278 (1966)

\bibitem{jbBBN} J.D. Barrow, Mon. Not. Roy. astr. Soc. \textbf{175}, 359
(1976)

\bibitem{BJS} J.D. Barrow, R. Juszkiewicz and D.N. Sonoda, Mon. Not. Roy.
astron. Soc. \textbf{213}, 917 (1985)

\bibitem{cmb} D. Saadeh, S.M. Feeney, A. Pontzen, H.V. Peiris and J.D.
McEwen, Phys. Rev. Lett. \textbf{117}, 131302 (2016)

\bibitem{midd} J. Middleton, Class. Quantum Grav. \textbf{27,} 225013 (2010)

\bibitem{vac1} M.E. Rodrigues and M.J. Houndjo, D.\ Saez-Gomez and F.
Rahaman, Phys. Rev. D \textbf{86,} 104056 (2012)

\bibitem{vac2} G.G.L. Nashed, Eur. Phys. J. Plus, \textbf{129,} 188 (2014)

\bibitem{ftSing} A. Paliathanasis, J.D. Barrow and P.G.L. Leach, Phys. Rev.
D \textbf{94,} 023525 (2016)

\bibitem{maluf} J.W. Maluf, Annalen der Physik \textbf{525,} 339 (2013)

\bibitem{aldro} R. Aldrovandi, J.G. Pereira and K.H. Vu, Braz. J. Phys. 
\textbf{34,} 1374 (2004)

\bibitem{tgr} S.G. Turyshev, Ann. Rev. Nucl. Part. Sci. \textbf{58,} 207
(2008)

\bibitem{tgr2} B.P. Abbott et al. (LIGO Scientific and Virgo
Collaborations), Phys. Rev. Lett. \textbf{116, }221101 (2016)

\bibitem{bamex} K. Bamba, C.Q. Geng, C.C. Lee and L.W. Luo, JCAP \textbf{1101%
}, 021 (2011)

\bibitem{Iorio2} L. Iorio, N. Radicella and M.L. Ruggiero, JCAP \textbf{2015,%
} 08 (2015)

\bibitem{Bas} S. Basilakos, Phys.\ Rev. D \textbf{93,} 083007 (2016)

\bibitem{Farrugia:2016xcw} G.~Farrugia, J.~L.~Said and M.~L.~Ruggiero,
Phys.\ Rev.\ D \textbf{93},104034 (2016)

\bibitem{rfa} R. Ferraro and F. Fiorini, Phys. Rev. D \textbf{84,} 083518
(2011)
\end{thebibliography}
\end{document}